\pdfoutput=1
\documentclass{article}
\usepackage{epsf,graphicx}
\usepackage[hidelinks]{hyperref}
\hypersetup{pdftitle={Certificates of Immortality for Gliders on the Hat and Spectre Tilings}, pdfauthor={James Edmond}}
\usepackage{amsmath}
\usepackage{color}
\usepackage{amssymb,ComplexSystems}

\newif\ifpreprint
\preprinttrue
\makeatletter
\ifpreprint
  \imagefalse
  \renewcommand{\@firstcopyrightfoot}{\hfil{\scriptsize Preprint.}\hfil}
  \def\@evenfoot{{\scriptsize Preprint}\hfil}
  \def\@oddfoot{\hfil{\scriptsize Preprint}}
  \def\@evenhead{\footnotesize{\sffamily\bfseries\thepage}\hfil\slshape Certificates of Immortality}
  \def\@oddhead{\footnotesize{\slshape Certificates of Immortality}\hfil{\sffamily\bfseries\thepage}}
\fi
\makeatother
\newenvironment{keywords}%
  {\par\vskip6pt\small\leftskip=2pc\noindent\textbf{Keywords:}\ }%
  {\par}
\newif\ifdraftnotes
\draftnotestrue

\newcommand{\Z}{\mathbb{Z}}

\newcommand{\gold}{\varphi}
\newcommand{\zt}{\zeta}
\newcommand{\lam}{\lambda}
\newcommand{\dg}{{}^{\circ}}
\DeclareMathOperator{\head}{arg}

\begin{document}
\title{Certificates of Immortality for Gliders\\
on the Hat and Spectre Tilings}
\author{\authname{James Edmond}\\[2pt]
\authadd{London, United Kingdom}\\
\authadd{james@offlattice.org}}
\markboth{Complex Systems}
{Certificates of Immortality}
\maketitle
\begin{abstract}
Cellular automata on the hat and spectre monotilings support gliders: localized patterns that
travel at constant speed and heading on a tiling that has no translational symmetry. An earlier paper found
them by search and left open whether they travel forever; here it is settled by finite computation.
A certificate is a finite set of local situations, a pattern with a collar of terrain, closed under
the rule on every environment the tiling can present, the environments enumerated by the tilings'
substitution transducer. Seven closed sets, covering the earlier paper's gliders, prove that each
never dies on any tiling of its hull and holds its heading inside a rigorous interval. Three further
gliders, from a rule hand-designed in 2023, one found independently in 2026, and the author's spectre
rule moved to the hat, certify on the lane the hat gliders share, with the same cocycle. An
independent program re-derives each certificate.
\end{abstract}
\begin{keywords}
aperiodic monotile; hat tiling; spectre tiling; cellular automata; gliders; computer-assisted proof;
substitution tilings; finite-state transducers
\end{keywords}

\section{Introduction}
\label{sec:intro}

The hat and the spectre are aperiodic monotiles \cite{hat,spectre}: single shapes that tile the
plane, but never periodically. Our earlier paper
\cite{paper1} found cellular automata on both tilings whose small travelling objects crossed a
million tiles at constant speed and heading. It called them gliders on that evidence, noting that
evidence was not proof.

On a periodic grid the proof comes free: a glider that re-forms itself one step along faces the same
grid again, so it travels forever. A hat or spectre tiling offers no such repeat. At every step a
glider meets terrain it may have met before, but the tiling as a whole is never the same twice, and
surviving so far says nothing about the step to come. How, then, can a glider on such a tiling be
proved immortal?

The answer lies in the locality of the rule. Each cell's next state depends only on its neighbors,
so whether a glider survives its next step depends only on its own cells and the tiles that
surround them. Although the tiling never repeats as a whole, its local patches do: within any
bounded radius only finitely many distinct patches occur, up to position.

A glider in flight therefore passes through a repertoire of \emph{situations}, each a pattern of active cells
with a \emph{collar} of surrounding tiles. Each situation has a finite set of possible successors:
the situations it can become after one step of the rule, over every legal way the tiling can
continue beyond the collar.

If a finite set of situations contains every possible successor of each of its members, and none
of them is the empty pattern, then a glider that enters the set can never leave it, and so it
never dies. We call such a set a \emph{certificate}.

We build certificates for seven objects under the earlier paper's four rules. Each one is immortal on every 
legal tiling (Theorem~\ref{thm:immortality}), and further, the certificates show that each keeps its 
orientation and advances at a speed bounded below, its heading inside a rigorous interval (Theorem~\ref{thm:lane}). 
The earlier paper defined a glider as a bounded pattern that travels
indefinitely; the certificate is finite, and the two theorems give the travel, so for these objects
the name is now a proven property. An independent program re-derives each certificate. Three
further hat gliders from outside our search certify by the same procedure: one under a rule
hand-designed by Trevorrow in 2023 \cite{hatlife}, one found by Gabbay's independent evolutionary
search \cite{aplife}, and one under our own spectre rule, which also flies the hat.
All three travel the same corridor of terrain as our hat gliders, with the same exact numbers.

The tiling never repeats globally, yet whether a glider flies forever is decided by a computation that
terminates, and its termination is part of the proof.

The contributions are as follows.
\begin{enumerate}
\item Immortality certificates for every glider of the earlier paper: seven closed sets, each
verified by an independent program (Section~\ref{sec:certificates}).
\item Three corrections to the earlier paper: one certificate per object, not six per tiling; its
``lane capture'' a distinct glider of the same rule (Section~\ref{sec:seven}); and a second lane
family on the spectre, $9.156\dg$ from the first (Section~\ref{sec:cocycle}).
\item Certificates for three gliders from outside the earlier paper's search, with the identity of
their cocycle numbers with the earlier paper's hat gliders' (Section~\ref{sec:outside}).
\item The controls: observed-transition, death-detection, seeding, collar-radius and
independent-checker tests, each stated before its result was known (Section~\ref{sec:seven}).
\end{enumerate}

Three devices make the construction work. Completeness of the terrain is the crux: Tatham's
substitution transducers \cite{tatham} read a tile's address one symbol at a time, and exploring
them enumerates every legal patch around a tile, with a termination argument that is the
completeness proof (Section~\ref{sec:atlas}). Closure is then checked over every way the
surrounding tiles can continue (Section~\ref{sec:closure}). Each transition carries an exact rigid
motion, and summed around the cycles of the certificate these motions bound the heading and the
speed (Section~\ref{sec:cocycle}).

Section~\ref{sec:prelim} fixes notation. Section~\ref{sec:certificates} builds the certificates.
Section~\ref{sec:outside} certifies the gliders from outside the earlier paper's search.
Section~\ref{sec:discussion} discusses what the shared corridor means and the limits of the method,
and Section~\ref{sec:repro} records the artifacts. The mechanism behind the shared corridor, the
exact value of the compass offset and the rhythm of the return times are the subject of a paper in
preparation.

\section{Preliminaries}
\label{sec:prelim}

\subsection{The tilings, their hierarchies, and addresses}
\label{sec:prelim-tilings}

The hat is the polykite $\mathrm{Tile}(1,\sqrt3)$ discovered by Smith, Myers, Kaplan and
Goodman-Strauss \cite{hat}, and a hat tiling uses both the hat and its mirror image, the anti-hat, the
latter with density $1/\gold^{4}$ of the former, $\gold=(1+\sqrt5)/2$. Every hat tiling decomposes uniquely into four \emph{metatiles}, clusters of
four, one, two and two hats called $H$, $T$, $P$ and $F$ (the $H$ carries the single anti-hat), and the
metatiles decompose in turn into \emph{supertiles} of every level: the substitution
replaces an $H$ by ten metatiles, a $T$ by one, a $P$ by five and an $F$ by six, and the level-$n$
supertiles of a tiling are the images of its metatiles under $n-1$ applications of it. Uniqueness of the
decomposition is the aperiodicity proof \cite{hat}; it makes the hat tilings the \emph{hull} of the
substitution, the set of tilings every finite patch of which occurs inside some supertile, and every
statement of this paper about ``every tiling'' is a statement about that hull.
\begin{figure}[t]
\centering
\includegraphics[width=\linewidth]{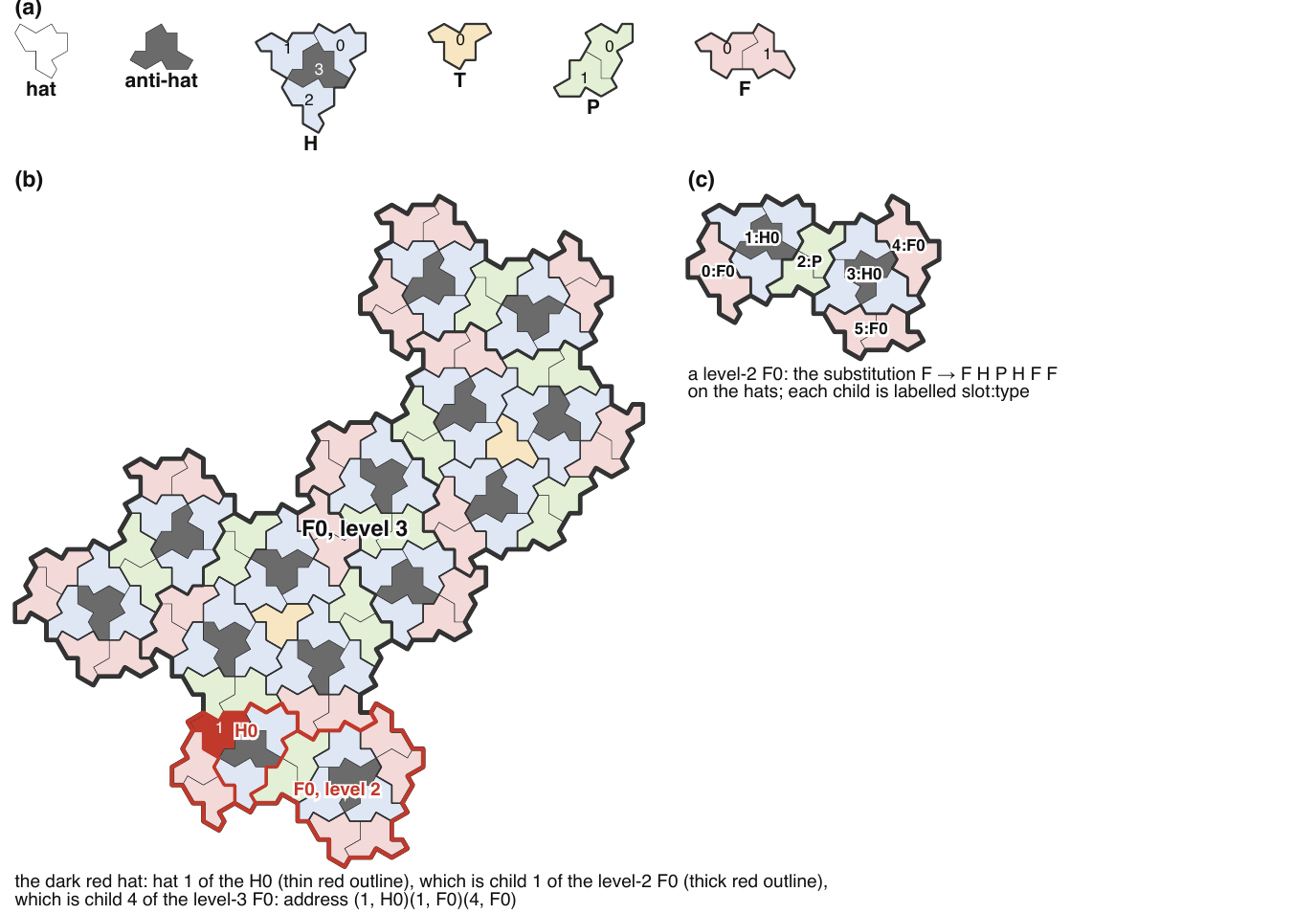}
\caption{The hat's hierarchy. (a) The hat, its mirror image the anti-hat, and the four metatiles $H$, $T$,
$P$, $F$ as clusters of four, one, two and two hats, each hat numbered by its slot; the anti-hat (dark)
is slot 3 of the $H$. (b) A level-3 $F_{0}$ supertile of 112 hats: level-1 supertiles filled by type, level-2
supertiles outlined, the level-3 outline heaviest. Every supertile is the image of a metatile under the
substitution. The dark red hat and its outlined ancestors illustrate an address: it is hat 1 of an $H_{0}$ (thin
red outline), which is child 1 of a level-2 $F_{0}$ (thick red outline), which is child 4 of the level-3
$F_{0}$, so its address is $(1,H_{0})(1,F_{0})(4,F_{0})$, read outward from the tile. (c) A level-2 $F_{0}$
supertile: the substitution $F\to FHPHFF$ carried out on the hats, each child metatile labelled by its
slot and type. The subscripts are the transducer's context variants of a metatile, which subdivide
identically; the polygonal metatiles of Refs.~\cite{hat,tatham} are the simplified outlines of these
clusters.}
\label{fig:hierarchy}
\end{figure}

\begin{figure}[t]
\centering
\includegraphics[width=\linewidth]{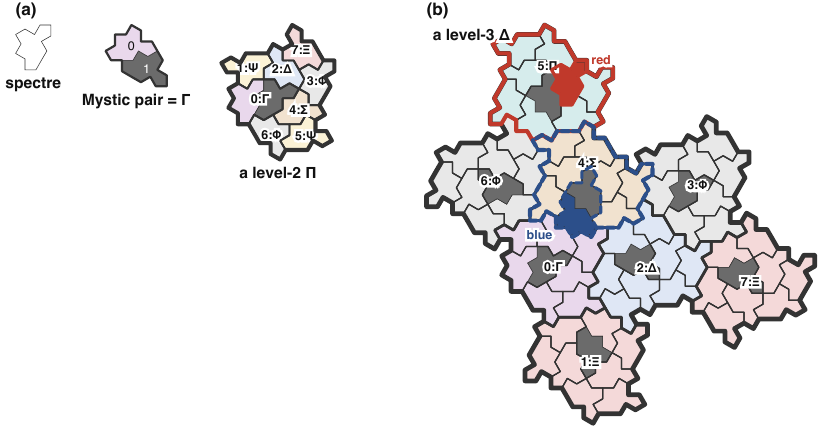}
\caption{The spectre's hierarchy. (a) The spectre; the Mystic pair, which is the metatile $\Gamma$,
two spectres of which the odd one, rotated by an odd multiple of $30\dg$ relative to the even tiles,
is dark; and a level-2 supertile of type $\Pi$ with its eight children labelled slot:type, the
substitution $\Pi\to\Gamma\,\Psi\,\Delta\,\Phi\,\Sigma\,\Psi\,\Phi\,\Xi$ on real tiles. The other
eight metatiles are single spectres, told apart by context. (b) A level-3 supertile of type $\Delta$,
its level-2 children filled by type and labelled, odd tiles dark. The red and blue tiles illustrate
two addresses: red is tile 0 of a $\Sigma$ (thin red outline),
child 4 of a level-2 $\Pi$ (thick red outline), child 5 of the $\Delta$, address
$(0,\Sigma)(4,\Pi)(5,\Delta)$; blue is tile 0 of a Mystic pair (thin dashed outline), child 0 of a
level-2 $\Sigma$ (thick dashed outline), child 4 of the $\Delta$, address $(0,\Gamma)(0,\Sigma)(4,\Delta)$.}
\label{fig:spectre-hierarchy}
\end{figure}

The spectre is the chiral monotile of the same authors \cite{spectre}: a single shape whose tilings use
no reflected copies, decomposing uniquely into nine metatiles, one of which, the Mystic, is a pair of
spectres and the other eight single spectres (Figure~\ref{fig:spectre-hierarchy}). Its substitution reflects: each step produces supertiles
of the opposite handedness, and only the square of the substitution is a subdivision in the usual
sense (Figure~\ref{fig:spectre-hierarchy} shows $\Pi\to\Gamma\,\Psi\,\Delta\,\Phi\,\Sigma\,\Psi\,\Phi\,\Xi$ on real tiles). The inflation factor per step is $\sqrt{\lam}$ with $\lam=4+\sqrt{15}$.

Tatham's transducers \cite{tatham} turn these hierarchies into a finite-state addressing scheme, and
everything computational in this paper runs on them. A tile's \emph{address} is the sequence of pairs
$(s_{1},T_{1})(s_{2},T_{2})\cdots$ recording that the tile is child (\emph{slot}) $s_{1}$ of a level-1 supertile of
type $T_{1}$, which is child $s_{2}$ of a level-2 supertile of type $T_{2}$, and so on; the types are
the metatiles refined by the context variants the transducer needs ($H_{0}$, $H_{1}$; $F_{0}$, $F_{1}$,
$F_{2}$; $P$; $T$ on the hat), and a tiling is fixed by giving one tile's address as an eventually
periodic sequence whose repeating part stays interior to its supertiles (our convention, following
Ref.~\cite{paper1}; Ref.~\cite{tatham} shows that the neighbors of eventually periodic addresses are
eventually periodic and computable). A \emph{patch} is the set of
tiles within a given graph distance of a root tile, generated by breadth-first search over the
\emph{neighbor transducer}: a finite-state machine (142 states on the hat, 150 on the spectre) that
reads a tile's address one symbol at a time and writes the address of its neighbor across a chosen
edge. The transducer is constructed and checked rather than derived from the tiles by a proof, and
Section~\ref{sec:certificates} uses it as a black box with four properties. Ref.~\cite{tatham} defines
the neighbor language $N$ of pairs of equal-length edge addresses (a tile address paired with one of
the tile's edges) whose tiles are adjacent inside the deflation, the subdivision into tiles, of a common
supertile, proves that $N$ is regular (its Theorem 1) and that each of its
sublanguages is zero-reversible (its Theorem 2), infers a recognizer for $N$ by Angluin's zero-reversible
state merging from a corpus of neighbor pairs computed by a recursive algorithm, checks that the
recognizer has a transition on every symbol of every legal edge address and enlarges the corpus until
it does (its Algorithm 5), and determinizes the result into a transducer with pending output. The
properties used are: (T1) soundness, every pair the recognizer accepts lies in $N$, which follows from
its Theorem 2 and Angluin's theorem; (T2) completeness, the recognizer accepts every pair in $N$,
which Ref.~\cite{tatham} asserts for a sufficiently large corpus and tests but does not certify; (T3)
bounded lag, the output trails the input by at most a fixed number of symbols (two, on both tilings),
so that a finite prefix of an address already determines the tile's neighbors; and (T4) accepting
states, entered once the recognizer has established a supertile common to the tile and its neighbor,
from which every further input symbol is copied to the output. (T2) is the one assumption about the
transducer that this paper does not inherit as a theorem; Section~\ref{sec:discussion} and
Appendix~\ref{app:crosscheck} give the evidence for it.

\subsection{The automata and the gliders}
\label{sec:prelim-ca}

The cellular automata act on the vertex-adjacency graph of a tiling: two tiles are neighbors if their
boundaries share a point, which gives every hat or spectre tile six or seven neighbors. Distances are
nevertheless measured in \emph{rings}, the graph distance under edge adjacency, the metric in which
patches are grown; a vertex-adjacency step crosses at most two rings on either tiling. Each tile
carries a state in $\{0,1,2,3\}$, with $0$ quiescent, and the rules are \emph{priority tables}: an
ordered list of \emph{clauses}, each specifying an own state (or a wildcard), a conjunction of thresholds on
the numbers $n_{i}$ of neighbors in state $i$, and a next state; the first matching clause fires, and a cell
matching no clause becomes quiescent. Conway's Life \cite{gardner} is the two-state case with the clauses (own 1,
$n_{1}\in\{2,3\}$, next 1) and (own 0, $n_{1}=3$, next 1). Transitions therefore depend on neighbor counts
but not on arrangement, as the variable neighborhood sizes require. The rules were found by evolutionary search
\cite{paper1}; their tables are reproduced in Appendix~\ref{app:rules} and every run of this paper is
a record --- tiling family, root address, radius, rule, initial states --- from which the run
regenerates exactly.

\begin{figure}[t]
\centering
\includegraphics[width=\linewidth]{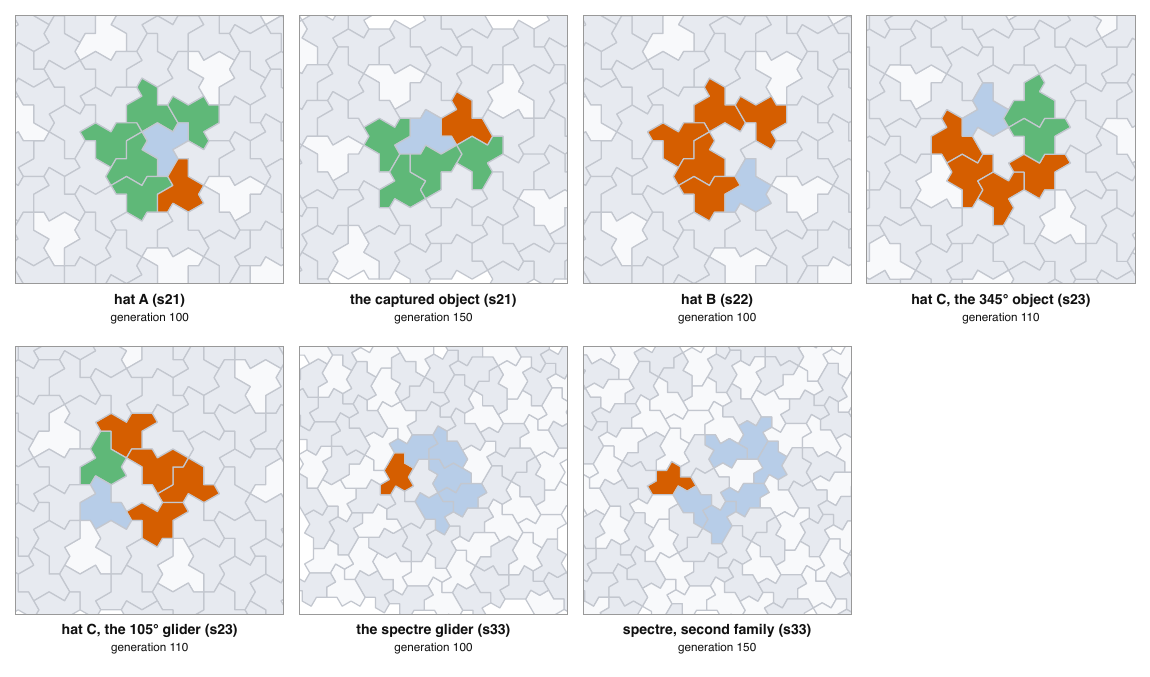}
\caption{The seven certified objects, one settled generation of each replayed from its record and drawn
at one scale: live cells in state 1 (orange), 2 (green) and 3 (pale blue); quiescent tiles pale, the
anti-hats white on the hat and the odd spectres darker on the spectre. Hat C's $345\dg$ object is one
member of the pair of Table~\ref{tab:certificates}.}
\label{fig:objects}
\end{figure}

Four rules and seven settled objects appear below; each object is a single glider in steady flight,
past the transient of its launch and whatever else the seed that first produced it launched alongside
(Section~\ref{sec:certificate-scope} makes ``settled'' precise as a component of the certificate graph). The \emph{hat A} glider (rule s21) is
the principal object: six to eight live cells in the three non-quiescent states, advancing one ring
every two generations. The same rule carries a second settled glider, the \emph{captured object},
which the earlier paper first met as a hat A flight that appeared to change lane
(Section~\ref{sec:seven}). Hat B (rule s22) is a different rule whose pattern of five to seven cells
on the same lane is hat A's with one cell removed and the states renamed
(Section~\ref{sec:seven}). Hat C (rule s23) carries two distinct objects, of six to nine and six to
ten cells. The \emph{spectre} glider (rule s33) is a head cell in state 1 with a wake of state-3
cells, four to six in all. The seventh object was found not by the earlier paper's search but by a
\emph{census}: for each of the four rules on its own tiling, every two-tile seed and a thousand random
six-cell seeds launched at sites spread across the patch, and the settled objects that resulted carried through the certificate
procedure below (Section~\ref{sec:certificate-scope} reports it). Under the spectre rule the census found,
besides the glider, four to eight cells on a second lane family about $9\dg$ from the glider's. Some seeds launch two copies of an
object together, a \emph{pair}; each member is treated separately below, and both close to the same
certificate. Figure~\ref{fig:objects} shows one settled generation of each. All advance at two
generations per ring; the earlier paper verified its objects by
cross-radius flights, the same launch replayed on patches of doubling radius, and, for hat A, hat B,
hat C and the spectre glider, by sliding-window flights of a million rings, in which a small patch
of tiles is regenerated ahead of the object as it moves and the frames of consecutive windows are composed
exactly.

A few words recur throughout. A \emph{flight} is a long run of the automaton; the million-ring flights
use a \emph{moving patch}, regenerated ahead of the object at each \emph{hop} and composed exactly with
the last, and we keep the earlier paper's name, sliding-window flight, for the method. A settled object's path is straight in the large;
the line it follows is its \emph{lane}, lanes carried into one another by the tiling's rotations form a
\emph{lane family}, and a family is named by the heading, in degrees, of the flight that first showed it,
in that record's frame, the frame fixed by the root tile's placement. The earlier paper's \emph{compass}
law states that the lanes are quantized to $60\dg$ with one measured offset per tiling, $45.523\dg$ on
the hat and $48.014\dg$ on the spectre; the six headings of a lane family are its \emph{spokes}. On the
hat the law stands; on the spectre the offset is one family's, and a second family runs about $9\dg$
from it (Section~\ref{sec:cocycle}). The sequence of terrains a lane presents to an object, as the
object's certificate records it in its situations and transitions (Section~\ref{sec:certificates}), is
the lane's \emph{corridor}. Along the lanes a ring is about
$4.3$ length units on the hat and $3.2$ on the spectre.

\subsection{Situations, collars, and keys}
\label{sec:prelim-situations}

The certificate's unit is the \emph{situation}: the local state of a glider together with as much of
the terrain as one step of the rule can see.

\begin{definition}[situation]
\label{def:situation}
The \emph{support} of a configuration is its set of non-quiescent cells, and its \emph{pattern} is the
states on the support. The \emph{$k$-collar} is the set of tiles within $k$ rings of the support. The
situation of collar radius $k$ is the pattern together with the tile classes (defined below) of the collar and the
\emph{gluings} among the support and collar tiles: which edge of which tile meets which edge of which
other. Its \emph{key} is the canonical encoding by breadth-first traversal in the tiles' intrinsic
edge order from a reference cell, minimized lexicographically over the choice of that cell in the support.
The \emph{centre} of a situation is a support cell minimizing the greatest ring distance to the
support, the first such cell in the key's traversal order when several tie, and that distance is its
support radius $r$.
\end{definition}
\begin{figure}[t]
\centering
\includegraphics[width=\linewidth]{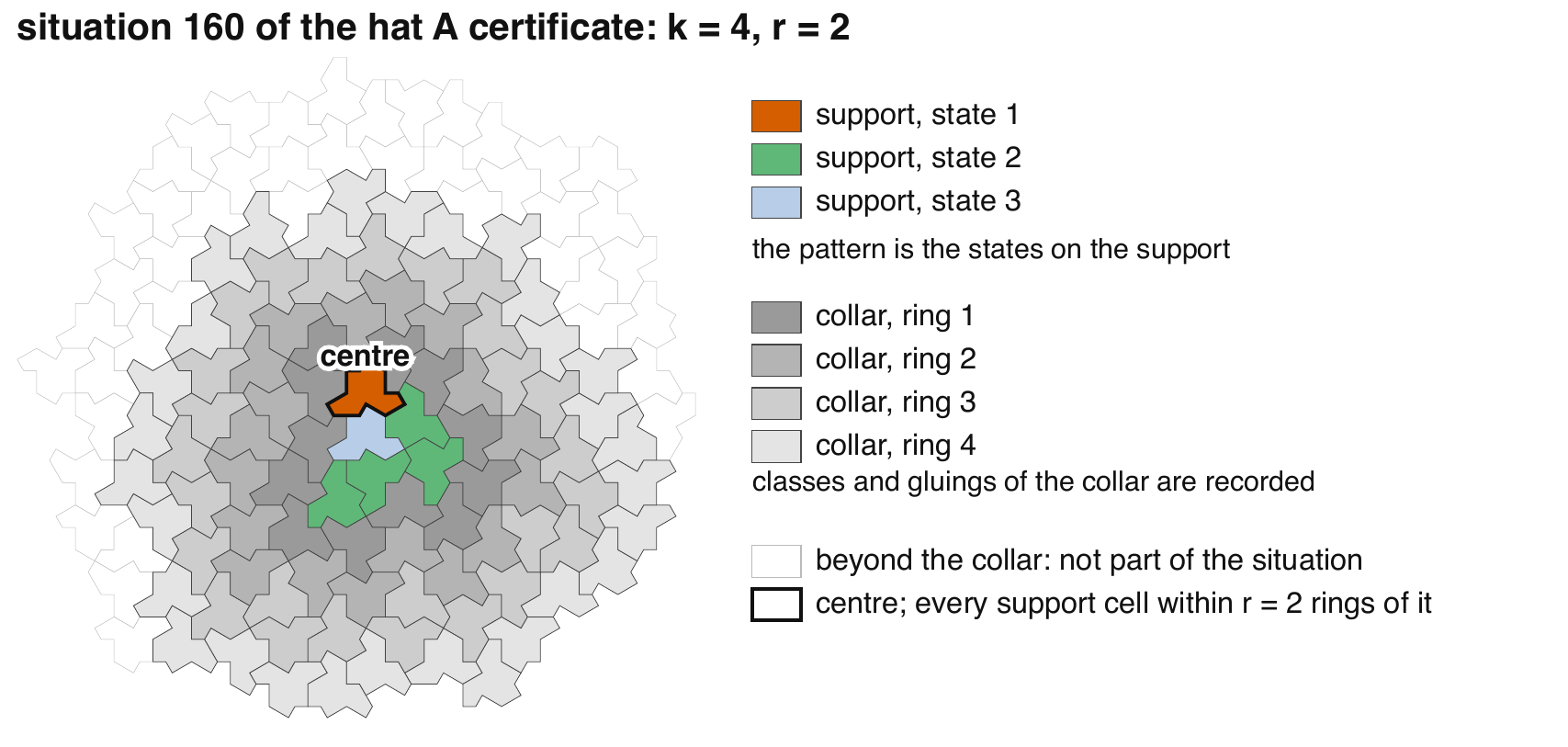}
\caption{The vocabulary of Definition~\ref{def:situation} on situation 160 of hat A's certificate.
The support is the six non-quiescent cells, filled by state, and the pattern is their states. The
collar is every tile within $k=4$ rings of the support, shaded by ring: 86 tiles, whose classes and
gluings the situation records; the tiles beyond it are not part of the situation. The centre is a
support cell from which every support cell lies within $r=2$ rings.}
\label{fig:anatomy}
\end{figure}
Figure~\ref{fig:anatomy} draws these terms on one situation.

\begin{figure}[t]
\centering
\includegraphics[width=\linewidth]{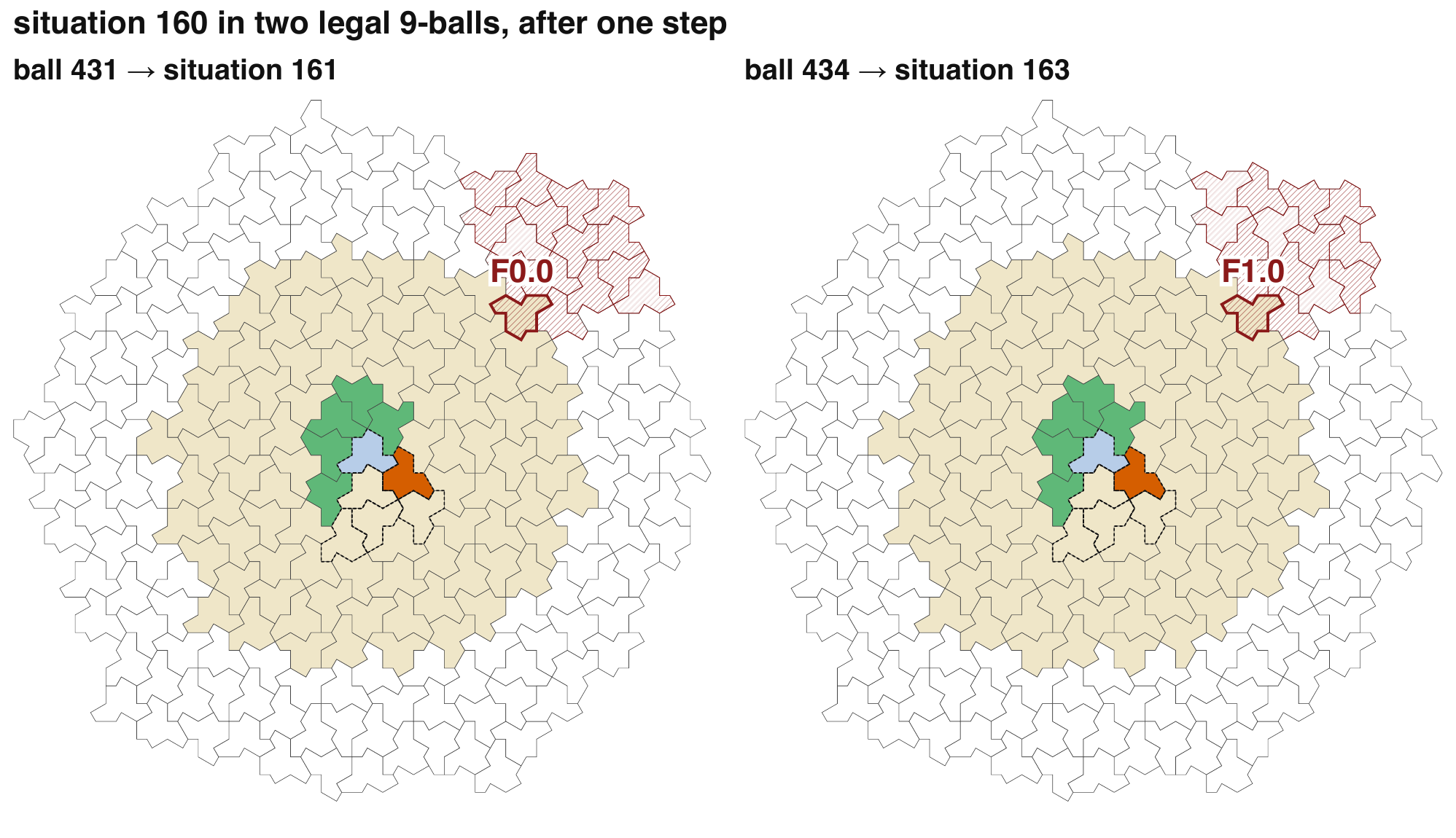}
\caption{Two legal extensions of a situation. Situation 160 of the hat A certificate
(Figure~\ref{fig:anatomy}) embedded in two legal 9-balls of the atlas, after one step of the rule. The
successor pattern is the same in both (filled; the old support dashed), but the successor's collar
(shaded) reaches one tile on which the two balls differ, and it differs only in its class label,
$F_{0}.0$ against $F_{1}.0$ (metatile variant and slot), the tile's shape and neighbors being identical: the step sees states and
adjacency alone, the certificate's keys record the classes, and the two balls lead to two different
successor situations, 161 and 163. The tiles on which the balls differ farther out, 35 and 36 of them, are
hatched. The
situation numbers are those of the certificate file.}
\label{fig:situation}
\end{figure}

The key records no coordinates and no absolute orientation, since edges are numbered intrinsically on
each prototile; a situation and its image under any of the tiling's rotations therefore have the same
key. Vertex degrees are not recorded either: the step is always computed on an exact patch, where they
are known. The collar radius $k=4$ is used throughout, which suffices for one step: a cell's vertex
neighbors lie within two rings of it, and their own neighborhoods within two more. The certified
gliders have $r=2$ or $3$. Figure~\ref{fig:situation} shows a situation and two of its legal extensions, atlas balls of
Section~\ref{sec:atlas} in which it embeds.

The \emph{class} of a tile is its label at the lowest level of the hierarchy, the metatile variant and
slot named by the first symbol of its address ($F_{0}.0$ is slot 0 of an $F_{0}$), read off when the patch
is generated. The rule cannot see classes: a step depends on states and adjacency alone. The certificate
records them because they carry the hierarchy into the situations and because closure with the finer
keys is the stronger statement;
closure holds equally on the class-blind quotient, the same closure with the classes stripped from
the keys (Table~\ref{tab:certificates}), which is all
Theorem~\ref{thm:immortality} needs.

\subsection{Notation and exactness}
\label{sec:prelim-notation}

Throughout, $\zt=e^{i\pi/6}$ is the primitive twelfth root of unity. Tile placements are exact: the
hat's edges have lengths $1$ and $\sqrt3$ in the twelve directions $\zt^{j}$, so the position of every hat
relative to the root tile is an element of $\Z[\zt^{2}]$, and the position of every spectre, whose edges are
unit edges in those directions, an element of $\Z[\zt]\cong\Z^{4}$, read off a walk along tile edges and
independent of the walk; both are recorded as integer vectors in the basis $1,\zt,\zt^{2},\zt^{3}$. The motions attached to the certificates'
transitions, and every quantity derived from them, are computed from exact sums in $\Z[\zt]$ and
converted to decimals only for printing; flight headings are floating-point, with error far below
the four decimals quoted.

\section{Certificates of immortality}
\label{sec:certificates}

A glider that has flown a million rings has met a million environments and survived them all; a
certificate must show that it survives every environment the tiling can present, including those no
flight has met. The argument has two layers. The first enumerates, completely, the legal
neighborhoods of a tile out to a fixed radius, using the substitution transducer in a way that makes
termination of the enumeration its own completeness proof. The second closes a finite set of
situations under the rule, stepping each situation inside every legal neighborhood that extends it.
A third step reads the exact displacement of every transition and bounds the heading.

\subsection{The legal-ball atlas}
\label{sec:atlas}

\begin{definition}[balls, atlas]
The \emph{$R$-ball} of a tile is the set of tiles within $R$ rings of it, with their classes and their
gluings; it determines the vertex adjacency among its tiles. A ball is \emph{legal} if it occurs in
some tiling of the hull. The \emph{atlas} of radius $R$ is the set of legal $R$-balls around a tile of
each base type (hat and anti-hat; spectre), each recorded with a \emph{witness}, a finite address
prefix that realizes it.
\end{definition}

The atlas is computed by exploring addresses through the neighbor transducer, one level at a time,
over \emph{configurations}: stored tiles with their types, classes and gluings, one live transducer
run per unresolved edge, copy links for runs that have accepted, and the type of the highest
supertile constrained so far. A configuration is a \emph{leaf} once every tile within $R$ rings of
the centre has a known class and no unresolved edge, and configurations are memoized by a key made
of everything just listed. Appendix~\ref{app:exploration} spells out the mechanics; the lemmas
below concern the algorithm as described there.

\begin{lemma}[merge soundness]
\label{lem:merge}
Two configurations with equal keys have identical futures: for every continuation of the address,
they reach the same leaves.
\end{lemma}
\begin{proof}
Feeding a symbol changes a configuration only through the transitions of its live runs on that
symbol, whose outputs are functions of the run's state and the symbol; through delivery of emitted
symbols to the identified neighbor, a tile named by the stored gluings; through duplication along copy
links; and through acceptance, a property of the run's state. Every ingredient is part of the key, and
the transducer's accepting states copy input to output, so no further state is needed for an accepted
run. Induction on the continuation gives the claim.
\end{proof}

\begin{lemma}[every legal ball is a leaf]
\label{lem:complete}
Let $B$ be an $R$-ball occurring in some tiling of the hull. Then $B$ is the ball of a leaf of the
exploration graph. Conversely, the ball of every leaf occurs in some tiling of the hull.
\end{lemma}
\begin{proof}
By the definition of the hull, $B$ occurs inside some level-$n$ supertile $T$. Its tiles then have
addresses of length $n+1$ ending in the symbol that names $T$, and any two of them that
share an edge are adjacent inside the deflation of $T$, so their edge addresses form a pair
of $N$ at that length; by (T2) the corresponding run has accepted by then, and by (T3) it has in any
case emitted the neighbor's type and edge, which is what the leaf condition needs. Hence the
configuration reached by a prefix of length at most $n+1$ of the centre's address is a leaf whose ball
is $B$. The exploration follows every legal symbol from every non-leaf configuration and merges only
configurations with equal keys, so by Lemma~\ref{lem:merge} the node reached by the prefix is in the
graph. Conversely, a leaf's witness is a legal finite address, and by (T1) every neighbor it names is
an adjacency inside the deflation of its top supertile, so its ball is realized in any tiling of the
hull containing that supertile.
\end{proof}

\begin{lemma}[termination]
\label{lem:termination}
The exploration terminates.
\end{lemma}
\begin{proof}
Keys are drawn from a finite set: the stored tiles lie in a fixed region, the transducer has finitely
many states, and copy links join stored tiles.
\end{proof}

Termination is thus the completeness proof. The hazard for any enumeration of balls is a corner
chain, a sequence of tiles lying near a supertile boundary at every level, whose balls are determined
only after many symbols; an enumeration of addresses with a length limit stops early on such chains
and misses their balls (the continuations left open in Appendix~\ref{app:crosscheck} are of this
kind). The exploration has no length limit to stop at. It follows configurations rather than
addresses, and since keys are finite, a chain of configurations that went on indefinitely would have to
revisit a key, at which point Lemma~\ref{lem:merge} says its future is already being explored from
the first visit; so the exploration closes every such chain into a cycle of the graph and, by
Lemma~\ref{lem:complete}, still reaches every leaf below it. That the program implements the
algorithm is checked independently.

\begin{proposition}[the checked graph; computer-verified]
\label{prop:graph}
For each atlas used below, the exploration graph is symbol-complete and faithful: every node's key
equals the recomputation from its address path alone, by the transducer library's own routine for the
neighbors of a finite address, iterated to a fixed point over the same region, with the state of each live run recomputed by
running the transducer over the tile's known prefix; every non-leaf's children are exactly the legal
parent symbols of its top type; every edge's target key equals the recomputation from the extended
path; and the leaves are exactly the configurations satisfying the leaf condition, with the recorded
balls.
\end{proposition}

\begin{figure}[t]
\centering
\includegraphics[width=0.8\linewidth]{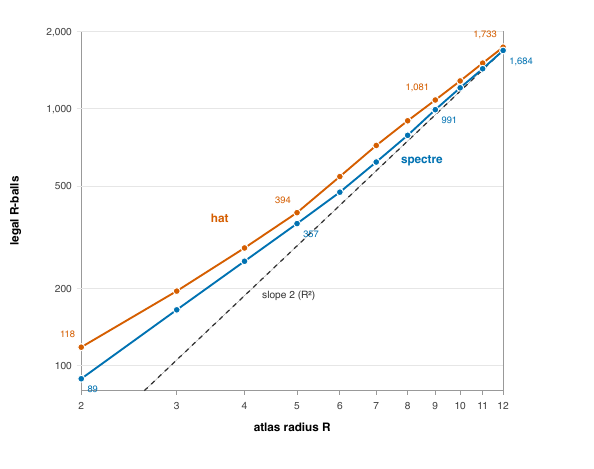}
\caption{The number of legal $R$-balls against $R$ on the two tilings (both base types on the hat),
with counts at four radii; all eleven counts, the leaf counts of the checked exploration graphs, are
tabulated in Appendix~\ref{app:exploration}. Log-log axes; the dashed guide has slope 2. The local
exponent rises with $R$, from about $1.2$ to $1.7$ on the hat and from $1.5$ to $1.9$ on the spectre,
towards the quadratic growth that linear repetitivity predicts.}
\label{fig:atlas}
\end{figure}

The check uses no incremental bookkeeping, no delivery queue and no copy links, so it isolates the
mechanisms in which an implementation error could hide. At $R=9$ it passes on the hat-centred graph
(5{,}004 nodes, 5{,}019 edges), the anti-hat-centred graph (612 nodes) and the spectre graph (5{,}374
nodes) with no mismatch of any kind, and likewise at the larger radii of Table~\ref{tab:certificates}:
$R=10$ on the hat (5{,}801 hat-centred and 810 anti-hat-centred nodes) and $R=11$ on the hat (6{,}911
and 929) and the spectre (6{,}699), every node, edge, leaf and ball agreeing. The $R=12$ atlases, used only
by the collar-radius control of Section~\ref{sec:seven}, are not covered by this check; no certificate
depends on them.
The atlases are also validated against the library's own patches
in both directions: every $R$-ball deep inside a radius-128 patch is in the atlas (46{,}303 of 46{,}303
on the hat at $R=9$; 44{,}688 of 44{,}688 on the spectre at $R=10$), with identical vertex adjacency, and
the patch realizes every ball of the atlas and no other, while every witness regenerates its ball.
With Lemmas~\ref{lem:merge}--\ref{lem:termination} and Proposition~\ref{prop:graph}, the leaves are
exactly the legal balls. Figure~\ref{fig:atlas} plots their number against $R$: it grows polynomially,
with an exponent rising towards the quadratic growth that linear repetitivity predicts
\cite{baakegrimm}, which these radii do not yet reach.

\subsection{Closure}
\label{sec:closure}

The step map on situations is computed as follows. For a situation $s$ with centre cell $c$, and for
every atlas ball $\beta$ whose centre tile has the base type of $c$, the situation is embedded in $\beta$
with $c$ at the centre: it is walked into the ball edge by edge from $c$, and the walk succeeds if
every base type, class and gluing agrees and every edge that leaves the situation's collar leaves its image.
Since the collar is a $k$-ball around the support and the support lies within $r$ of $c$, the
embedding is determined by the ball once $R\ge r+k$; the implementation asserts the stronger margins
it needs, that the support lies within $R-4$ of the ball's centre and that the successor's collar lies
within $R-1$, and $R$ is chosen per object accordingly ($R=9$ to $11$). The rule is then applied once
to the pattern on the ball's exact vertex adjacency, where every degree is known, and the successor
configuration, with its collar read from the ball, is canonicalized to a key. The successor of $s$ in
$\beta$ is that key, together with the rigid motion carrying $c$ to the successor's centre.

\begin{lemma}[closure is safe]
\label{lem:closure}
Let $S$ be a set of situations such that, for every $s\in S$ and every atlas ball $\beta$ in which $s$
embeds, the successor of $s$ in $\beta$ lies in $S$. Then on every tiling of the hull, a configuration
whose situation lies in $S$ has a successor whose situation lies in $S$.
\end{lemma}
\begin{proof}
The terrain around any occurrence of $s$ in a hull tiling is a legal $R$-ball around its centre,
which by Lemma~\ref{lem:complete} is in the atlas, and the rule's step on the tiling depends only on
that ball. A missing ball would break the argument; extra balls could only add transitions.
\end{proof}

\begin{theorem}[immortality; computer-verified]
\label{thm:immortality}
For each of the seven objects of Table~\ref{tab:certificates} there is a finite set of situations,
closed in the sense of Lemma~\ref{lem:closure}, containing no situation with empty support (death). On any
tiling of the hull, a configuration whose situation lies in the set has an evolution under the object's
rule that never dies.
\end{theorem}
\begin{proof}
By Lemma~\ref{lem:closure} the situation stays in the set forever, and no situation of the set is
empty. The sets themselves are the certificates, computed by the closure below and verified by the
independent checker of Section~\ref{sec:seven}. Like every computation of the paper, the theorem
inherits the transducer's completeness (T2) of Section~\ref{sec:prelim-tilings}, which is checked
rather than proved.
\end{proof}

The closure is computed by breadth-first search: the set is seeded with the situations observed on a
flight after its launch transient, the generations before the flight's situation first lies in the set
(hat A: every situation from generation 60 of the radius-384 record, 163 of them), and each situation is stepped in every ball it embeds in until no new key appears. For
hat A this discovers 218 further situations, for 381 in all, joined by 1{,}089 transitions using 529 of
the 1{,}081 balls; the closed set has no empty situation, and the search records no embedding
failure. Every certificate is a single strongly connected
component of its transition graph with no source state, so every situation recurs from every other on
some legal continuation, so no situation is transient. Branching, a
situation with several successors, occurs at 34 situations of hat A's certificate, where the terrain
ahead decides among two or three continuations; a single flight takes only a few of these branches, and
the closure follows all of them.

\subsection{The cocycle}
\label{sec:cocycle}

Every transition carries the exact rigid motion from the centre of a situation to the centre of its
successor: a rotation index, a multiple of $30\dg$, and a translation in $\Z[\zt]$, read from the
tiles' exact placements in the ball; this edge labelling is the certificate's \emph{cocycle}. Any
other canonical choice of centre would change each situation's
translations by a fixed offset of its own, but would leave every cycle sum unchanged. A breadth-first assignment of an orientation to every situation,
checked for consistency on every edge, decides whether any cycle of the graph rotates; then the
translations are rotated into a common frame, and Karp's minimum-cycle-mean algorithm \cite{karp}, run on exact
weights, bounds the mean displacement per generation of every cycle, its minimum and maximum along and
its minimum and maximum across the direction of a \emph{reference cycle}, the first cycle reached by
following each situation's first listed successor from the certificate's first situation; its direction
only fixes the along and across axes (on the hat it is a 26-generation cycle, Section~\ref{sec:seven}).

\begin{theorem}[lane; computer-verified]
\label{thm:lane}
On the closed set of each of the seven certified objects of Table~\ref{tab:certificates} the
orientation is constant: no cycle of the certificate graph rotates. Every cycle advances along the
reference direction by at least $2.1396$ length units per generation for the hat objects, $1.5807$ for
the spectre glider and $1.5441$ for the second spectre object. Every accumulation point of the
direction of a trajectory's displacement lies in an explicit interval of headings containing the
measured lane: of width $0.1189\dg$ for the hat objects, $0.3926\dg$ for the spectre glider, and
$0.5537\dg$ for the second spectre object.
\end{theorem}
\begin{proof}
A trajectory in the closed set is a walk on the certificate graph, and its displacement after $n$
generations is the sum of the edge translations, rotated by the walk's orientation, which is constant.
Any walk decomposes into simple cycles and a remainder of fewer than $|S|$ edges, so the mean
displacement per generation is a weighted average of the cycle means, each cycle's mean displacement per
generation weighted by the fraction of the walk's generations spent on it, up to the remainder's
$O(|S|/n)$: it lies within that distance of the convex hull of the cycle means, the set of all such
weighted averages, whose along and across components are bounded by the four extremes computed. The interval is the set of directions of vectors with those component bounds, and positivity
of the along minimum gives the speed bound.
\end{proof}

\begin{figure}[htbp]
\centering
\includegraphics[width=0.9\linewidth]{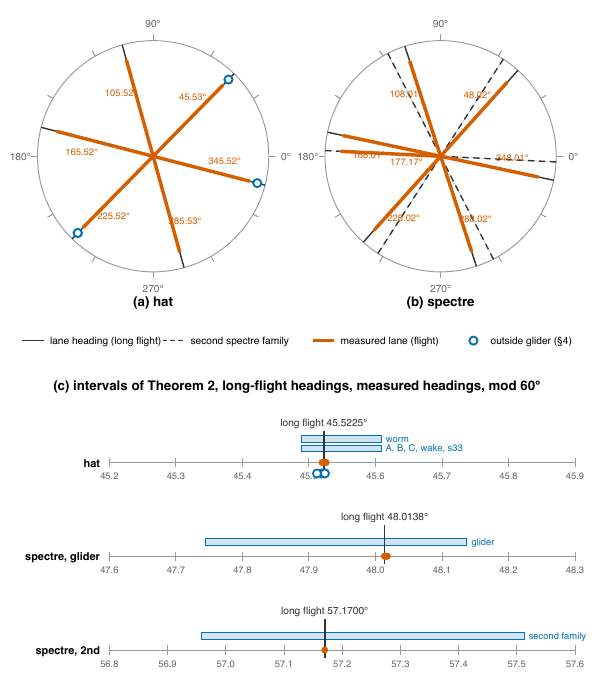}
\caption{Theorem~\ref{thm:lane} against the flights. (a) The hat: six spokes at hat A's
count-weighted return heading, $45.5225\dg+60k\dg$ (Section~\ref{sec:outside}); arrows, the six lanes
flown in the earlier paper, labelled with each flight's net-displacement heading; circles, the gliders
from outside the census (Section~\ref{sec:outside}). (b) The spectre: the glider family's spokes at the
heading of its million-generation flight, $48.0138\dg+60k\dg$, with its six flown lanes, and the
second family's at $57.1700\dg+60k\dg$ (dashed), the heading of its million-generation flight. (c) All of it modulo
$60\dg$: each certificate's rigorous heading interval (bars), the long flight's heading (line) and the
measured headings (dots; the error bar is the difference between a flight's whole and second-half
headings). Every measurement lies inside its interval, within a few thousandths of a degree of the long
flight, and the intervals are 30 to 100 times wider than the measurements' scatter.}
\label{fig:certcompass}
\end{figure}

The hat certificate has a further structure. Fix a spanning tree of the transition
graph and place every situation where the tree's motions put it, relative to one situation taken as
origin: the layout of Figure~\ref{fig:certificate}. Each non-tree edge then has a \emph{reduced displacement}, the vector
from where its motion lands to its target's tree position; tree edges contribute nothing, so the
displacement of any cycle is the sum of the reduced displacements of the non-tree edges it uses. Hat
A's 709 non-tree edges carry only four distinct reduced displacements: zero (the motion lands exactly on
its target), two vectors $A=16+46\zt^{2}$ and $B=10+28\zt^{2}$ in the integer coordinates of
Section~\ref{sec:prelim-notation} (in the plane $A=(39,23\sqrt3)$ and $B=(24,14\sqrt3)$, since
$\zt^{2}=e^{i\pi/3}$), and their sum $A+B$; the tree of Figure~\ref{fig:certificate} adds $2A+B$. We
read the vectors off the certificate: $A$ and $B$ are the two shortest non-zero values, and the reference
cycle of 26 generations is $A$ itself. So every cycle's displacement is $mA+nB$ with non-negative
integers, and every long-run heading is $\head(A+\varrho B)$ with $\varrho$ the ratio of
$B$-traversals to $A$-traversals. The five hat objects
have identical cocycle numbers, including the same 26-generation reference cycle of displacement
55.749 units, and differ only in the states they carry through the corridor.
The spectre certificates carry six or seven reduced vectors within a few degrees of one another. In
neither case are the cycles collinear, so the exact heading is not a property of any cycle but the
ergodic average of the cycles a hull tiling visits, which is what the long flights measure.

Theorem~\ref{thm:lane} also corrects the earlier paper's compass on the spectre, whose single offset per
tiling was one object's. Every heading the theorem allows the second spectre object lies between
$8.8\dg$ and $9.8\dg$ from every heading it allows the spectre glider (Table~\ref{tab:certificates}),
so the two lane families are distinct, and the million-generation flights of the two put the
separation at $9.156\dg$.

\subsection{The seven certificates}
\label{sec:seven}

\begin{table}[t]
\caption{Certified objects. $R$ is the atlas radius and $k=4$ the collar radius throughout;
``observed'' counts the situations seen on the seeding flight, ``closed'' the closed set, ``blind'' its
quotient with tile classes stripped; the heading interval is that of Theorem~\ref{thm:lane}, quoted
modulo $30\dg$ because tile orientations on both tilings are multiples of $30\dg$, so the cocycle's
rotation index is too; ``171 / 164'' are the two
members of the pair. The pair objects share a certificate (identical closed sets from either member). The second
spectre object was closed from a census launch rather than a flight.}
\label{tab:certificates}
\begin{center}
\setlength{\tabcolsep}{3pt}\scriptsize
\begin{tabular}{@{}llrrrrrl@{}}
\hline
object & rule & $R$ & obs. & closed & blind & trans. & heading (mod $30\dg$)\\
\hline
hat A & s21 & 9 & 163 & 381 & 271 & 1{,}089 & $[15.4895\dg,\,15.6084\dg]$\\
hat B & s22 & 10 & 163 & 381 & 271 & 1{,}340 & same\\
captured object (s21) & s21 & 9 & 121 & 630 & 352 & 1{,}621 & same\\
hat C, the $345\dg$ pair & s23 & 11 & 171 / 164 & 381 & 275 & 1{,}582 & same\\
hat C, the $105\dg$ glider & s23 & 11 & 146 & 257 & 191 & 801 & same\\
spectre glider (pair) & s33 & 9 & 306 / 353 & 2{,}542 & 1{,}920 & 7{,}089 & $[17.7448\dg,\,18.1374\dg]$\\
spectre, 2nd family & s33 & 11 & --- & 1{,}454 & 1{,}110 & 5{,}675 & $[26.9589\dg,\,27.5126\dg]$\\
\hline
\end{tabular}
\end{center}
\end{table}

\begin{figure}[t]
\centering
\includegraphics[width=\linewidth]{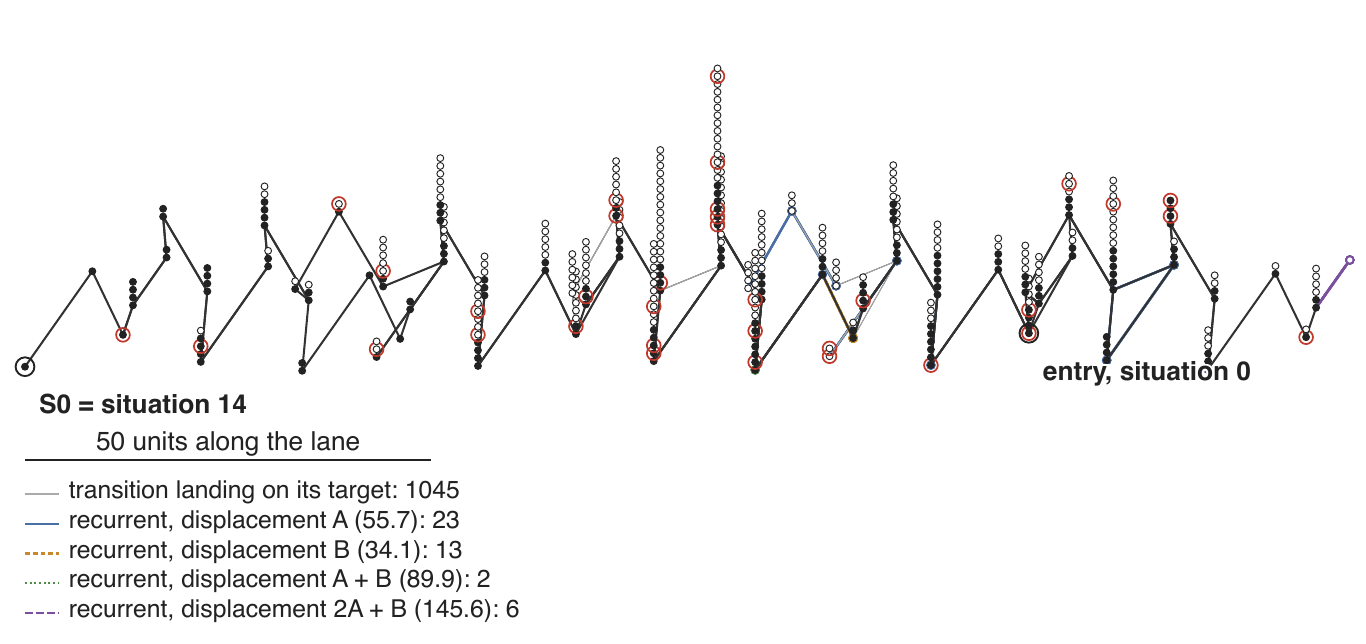}
\caption{Hat A's certificate unrolled along its lane. Each of the 381 situations is placed where its
centre lies relative to the situation visited most often in flight, taken as origin, following a
spanning tree of the transition graph with the exact motions of Section~\ref{sec:cocycle}; the lane is horizontal and the perpendicular
offset is exaggerated threefold. Situations sharing a position, which differ only in their collars,
are stacked as beads: 381 situations occupy 61 positions, up to 21 at one of them. Filled beads were
observed on the seeding flight, open ones were found only by the closure; the flight's 167 transitions
are heavy, the other 922 thin; the 34 branching situations are ringed. A transition whose motion does
not land on its target's tree position is a recurrent edge, coloured by the displacement between
landing point and target, an integer combination of the cocycle generators $A$ and $B$ (here $A$, $B$,
$A+B$ and $2A+B$); every cycle of the graph is built from these, and the certificate is one strongly
connected component.}
\label{fig:certificate}
\end{figure}

Table~\ref{tab:certificates} lists the certificates, and Figure~\ref{fig:certificate} draws hat A's
along its lane; Figure~\ref{fig:certcompass} sets their heading intervals beside the flights. Each closed with no death and no embedding
error, each is one strongly connected component, and on each the orientation is locked (constant, Theorem~\ref{thm:lane}). Two remarks
on what is counted. First, the situation key is blind to rotation, and the hull is invariant under the
rotations by multiples of $60\dg$ because a rotated tiling by the same tile is again a tiling by it, so
a settled object on one lane and its rotated copy on another have the same keys:
one closed set covers all six lanes of an object, and the earlier paper's ``six certificates per
substrate'' are one per settled object. Second, the objects are distinguished by their
sets: the captured object of rule s21 (the earlier paper's lane capture) shares no situation with hat A, with or without classes, and is
a different settled glider of the same rule, not hat A on another lane; likewise hat C's rule carries
two distinct settled gliders. Hat A and hat B, under different rules, share no situation while having
the same number of situations and the same cocycle numbers, and the sets explain why: they are the
same mechanism in two state alphabets. Renaming hat A's states 1 and 2 to 3 and 1 and deleting its state-3
cell re-encodes its 381 situations onto hat B's 381 one to one and carries its 1{,}089 transitions
onto transitions of hat B (the identity map, and the renaming without the deletion, carry none), and
on one patch the two gliders occupy the same tiles at the same generations for 381 generations, hat A
with one extra cell in state 3 that hat B's rule does without. The two objects are distinct, by that
cell and by their rules.

The following controls were run, each stated before its result was known.
\begin{enumerate}
\item \emph{Observed transitions.} Every consecutive pair of situations on the seeding flight is a
transition of the certificate with the same rigid motion (hat A: 672 of 672; every object likewise,
the second spectre object on the census launch that seeded it),
and every centre ball met on the flight is in the atlas.
\item \emph{Death is detectable.} Stepping hat A's situations with hat B's rule does not certify: the
closure finds death transitions and 2{,}476 states with populations up to 15.
\item \emph{Seeding.} Closure from the single situation at generation 300 of hat A's flight yields the
identical 381-state set.
\item \emph{Collar radius.} Every object also closes at $k=2$, $5$ and $6$ (with atlases to $R=12$),
and projecting a finer closure onto a coarser collar reproduces the coarser closure, with two
exceptions. The captured object's $k=4$ set of 630 situations is the union of three
distinct gliders of hat A's rule, of 257, 256 and 127 situations, whose patterns largely overlap; at
ten situations two of them share the four-ring collar as well, so the situation is the same for both,
and the closure, which sees nothing beyond the collar, must follow the legal extensions of both and
walks from one glider's repertoire into the other's. With five rings no situation is shared and each
glider closes on its own. Hat C's $105\dg$ object shows the same gluing at $k=2$.
Theorem~\ref{thm:immortality} is unaffected, since a closed union is closed and death-free; what the
collar radius decides is object identity, which is therefore read from the stabilized sets.
\item \emph{An independent checker.} A program written from the file formats and the rule table
alone, sharing no code with the implementation and taking the atlas file as data, decodes every
situation, embeds it in every ball of its centre's type, steps it, re-canonicalizes the successor, and checks that the set is closed, that
no successor is empty, that the listed transitions are exactly those found with the same motions,
and that an all-quiescent neighborhood stays quiescent at every vertex degree the atlas contains; it then recomputes
the cocycle. On every certificate of this paper, the seven of Table~\ref{tab:certificates} and the three
of Section~\ref{sec:outside}, it re-derives every transition, finds none unlisted, and reproduces every
cocycle number to the printed digit. It does not, and cannot, verify the transducer
or Lemmas~\ref{lem:merge}--\ref{lem:termination}; those rest on the transducer's properties
(T1)--(T4) of Section~\ref{sec:prelim-tilings}, of which (T2) is checked rather than proved
(Appendix~\ref{app:crosscheck}), on the faithfulness check of Proposition~\ref{prop:graph}, and on
the patch validation above.
\end{enumerate}

Each certificate also records its \emph{launch bridge}: the committed record, root and radius of the
seeding flight, and the generation at which the flight's situation first lies in the closed set (hat
A: generation 60 of its record, after which the flight remains in the set through its last generation
on the patch). The theorem therefore covers each committed launch from its bridge generation on; the transient
before it is a finite, recorded replay.

\subsection{What the theorem says, and does not}
\label{sec:certificate-scope}

Theorem~\ref{thm:immortality} certifies settled objects: once any evolution of the rule on any hull
tiling is in one of the certified situations it never dies, and by Theorem~\ref{thm:lane} it then
holds its orientation, advances at a bounded rate, and keeps its heading inside the interval. It says
nothing about other configurations of the same rule, and it is per object: the object is the strongly
connected component, and a rule may carry several. The census of Section~\ref{sec:prelim-ca}, launching every
two-tile seed and a thousand random six-cell seeds at sites spread across the patch, found the following
objects, each certified by the same procedure. Each hat rule carries several single gliders beyond
those of Table~\ref{tab:certificates}, all on the same lane family, with the same speed bound and heading
intervals within $0.0003\dg$ of theirs. Hat A's and hat B's rules also carry formation pairs (two single
gliders travelling side by side); hat A's carries bound pairs (two single gliders travelling locked
together) and hat B's carries gliders with a passenger (a single glider with a few extra cells attached).
The spectre rule carries the glider of Table~\ref{tab:certificates}, formation pairs of it, and the second
object, on a second lane family. Each of these objects can be replayed at
\href{https://offlattice.org/monotile/census/}{offlattice.org/monotile/census/}, and the repository holds its
record and the command that rebuilds its certificate. The theorem is also silent on the exact heading.

\section{Gliders from outside the census}
\label{sec:outside}

\begin{figure}[t]
\centering
\includegraphics[width=0.78\linewidth]{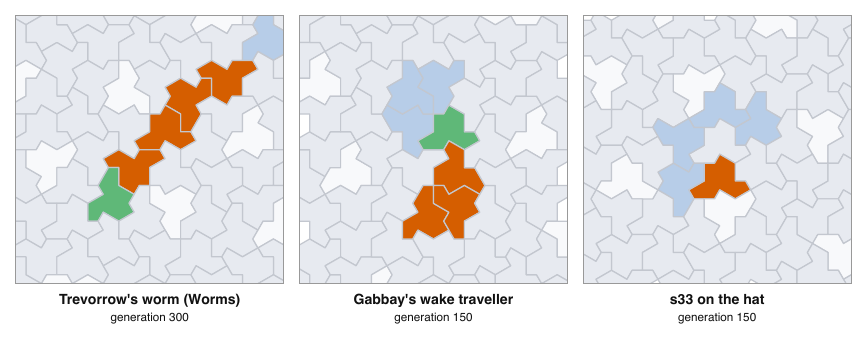}
\caption{The three gliders from outside the census, one settled generation of each at the scale of
Figure~\ref{fig:objects}: Trevorrow's worm (states 1 body, 2 head, 3 tail), Gabbay's wake traveller,
and the spectre rule s33 on the hat.}
\label{fig:outside}
\end{figure}

The certificates of Section~\ref{sec:certificates} were built for the objects our own search produced,
and one may ask whether that search selects for objects of this kind. Three hat gliders under rules
from outside our search allow a test, and each answers a different part of the question. Trevorrow's worm was designed by hand to run along the gaps between rows of
reflected hats, so its direction was chosen and its heading is not a test; what it tests is that a
glider with no search behind it is immortal by the same certificate, and that the corridor it runs in
is hat A's. Gabbay's traveller came from an evolutionary search that was not ours and, like ours,
rewarded distance and compactness and did not specify a direction, so it tests the direction as well. Our
spectre rule moved to the hat tests whether the corridor depends on the rule or on the tiling the rule
was found on. Figure~\ref{fig:outside} shows the three. Each was run through the pipeline unchanged: the record format, the atlas of Table~\ref{tab:certificates}, the closure, the
cocycle, and a \emph{keyed flight} of $20{,}000$ generations, in which the situation key is computed at
every generation and looked up in the certificate. Two notions are needed to compare them. The
\emph{return words} of a certified object are the stretches of a keyed flight between consecutive
visits to its most-visited situation, named by their lengths in generations; their displacements are
integer combinations of the cocycle generators $A$ and $B$ of Section~\ref{sec:cocycle}, quoted here
rotated into the lane's frame, so that a lane running opposite to $A$ in the root frame, as hat A's
does, shows negative combinations; for every hat object whose return words are reported here they are
consecutive terms of one additive sequence $26$, $42$, $68$, $110$, $178$, \ldots, each term the sum of
the two before. And the \emph{count-weighted heading} of an object is the direction of the sum of its
return displacements weighted by their counts; for hat A it is $225.522493\dg$ over $10^{6}$
generations, and it serves below as the lane's reference heading.

\subsection{A glider designed by hand: Trevorrow's worm}
\label{sec:worm}

After the first version of the earlier paper was posted, Andrew Trevorrow pointed out his
\emph{HatLife} \cite{hatlife}: an interactive environment announced in May 2023, the year the hat was
found, that runs a selection of cellular automaton rules on a fixed patch of 21{,}781 hats. Among
HatLife's rules is \emph{Worms}, four states on the vertex neighborhood (0 empty, 1 body, 2 head, 3 tail), designed so that short chains travel the
wide gaps between rows of reflected hats (anti-hats), which the rule holds permanently empty; in the
thin gaps, its author notes, they do not survive for long. Trevorrow's fixed patch could show the worms
travelling but not whether they travel forever, and the rule was written around the reflected-hat gaps. The gaps, however,
run along the lane directions. On the
radius-512 patch, a line through an interior reflected hat along any of the six compass directions
carries a median of 130 to 180 reflected hats within $0.8$ units of it, at spacings that take only
four or five distinct values on a typical line, among them $21.6$, $34.1$ and $55.7$ units, the
lengths $|A|-|B|$, $|B|$ and $|A|$ of Section~\ref{sec:cocycle}; a line $17\dg$ off a spoke carries a
median of 57, at spacings that are all different. The reflected hats form straight rows along the lane
directions, the one class of tile whose rows the eye picks out, and a rule built to run between two
such rows is built to run along a lane. The worm's heading is therefore no test of the compass. The
test is its certificate, whose cocycle is hat A's.

The rule was transcribed verbatim into the machinery of this paper as a stratified priority table, one table per
chirality, hats carrying the Worms table and anti-hats a table with no rows (Appendix~\ref{app:rules});
the transcription was checked against a direct implementation of HatLife's transition function for 400
generations on a random soup, and the two tilings against each other through their vertex degrees, reflected
hats 6 and unreflected 6 or 7 in both. Worms arise from random soups as Trevorrow reported. One of
them, of population eight and extent five (rings spanned), seeded from a random soup on the hat root of the earlier
paper's records (the record is committed), was run through the pipeline unchanged: the radius-9 atlas
and collar radius 4 of Table~\ref{tab:certificates}, closure from its keyed flight on a radius-384
patch (585 generations from generation 150, 192 observed situations, population 6 to 8), the exact
cocycle, and a keyed flight of 20{,}000 generations. The closure is 326 situations and 644
transitions, closed with no death and one strongly connected component, orientation locked; the flight
has no generation off the certificate, uses 195 of its transitions, and its exact sum of certificate
motions matches the geometric displacement of $4.28\cdot10^{4}$ units to $10^{-9}$. Theorem
\ref{thm:immortality} applies: the worm is immortal on every tiling of the hull.

\begin{table}[t]
\caption{Trevorrow's worm against hat A. The worm's certificate is its own, closed from its own flight
at the same $R$ and $k$, and the quantities below are read from it exactly as from hat A's. The return
words are those of each object's most-visited situation (Section~\ref{sec:outside}); the worm's returns
are the next two terms of the same additive sequence. Headings in the frame of the shared root.}
\label{tab:worm}
\begin{center}
\setlength{\tabcolsep}{3pt}\scriptsize
\begin{tabular}{@{}lll@{}}
\hline
 & Trevorrow's worm (2023) & hat A (s21)\\
\hline
observed / closed situations & 192 / 326 & 163 / 381\\
transitions & 644 & 1{,}089\\
cocycle generators & $A$, $B$ & $A$, $B$\\
speed bound (units per generation) & $2.1396$ & $2.1396$\\
heading interval (mod $30\dg$) & $[15.4895\dg,\,15.6087\dg]$ & $[15.4895\dg,\,15.6084\dg]$\\
return times (generations) & 68, 110 & 42, 68, 110\\
return displacements & $-(2A+B)$, $-(3A+2B)$ & $-(A+B)$, $-(2A+B)$,\\
 & & $-(3A+2B)$\\
count-weighted heading & $225.522495\dg$ & $225.522493\dg$\\
\hline
\end{tabular}
\end{center}
\end{table}

Table~\ref{tab:worm} sets the worm's certificate beside hat A's. The two objects have different states
and different authors and share no situation, yet every number that belongs to the corridor agrees: the same
generators $A$ and $B$, the same speed bound, heading intervals agreeing to within $0.0003\dg$, return
displacements from the same lattice $-(mA+nB)$ at consecutive terms of the same additive sequence,
and a count-weighted heading within two millionths of a degree of hat A's. The second worm of the same soup, of population seven,
flies the opposite spoke, $45.519\dg$ against the compass $45.5225\dg$ on a radius-384 flight. Hat A
and hat B were the earlier paper's evidence that the corridor is a property of the
tiling; they came from one search, and Section~\ref{sec:seven} finds them to be the same mechanism
in two alphabets. The worm is a different object, on a parallel line about twenty units from hat A's path
and sharing no tile with it, under a rule written by hand, for the reflected-hat gaps, before the
corridor was known to exist.

\subsection{A glider from an independent search: Gabbay's wake traveller}
\label{sec:wake}

Murdoch J. Gabbay's \emph{Aperiodic Life} \cite{aplife}, released in October 2026, is a browser cellular
automaton on square, hexagonal, hat and Penrose boards, its hat patches generated independently from
Kaplan's substitution with the edge-or-corner adjacency, our vertex adjacency. It reproduces the earlier paper's hat gliders
at placements of its own choosing, with three implementations agreeing, and its evolutionary search,
randomly initialized and sharing no code with ours, found a four-state priority table of four rows (Appendix~\ref{app:rules}), under which a six-cell seed flew $1{,}000$ generations
on his $372{,}100$-tile patch and which is not, he checks, a relabelling of any of our rules.
Transcribed as data and run on our patches, the rule ignites compact
travellers from every random soup; one of them keeps population 4 to 8 and extent 2 to 3 on patches of
radius 48, 96, 192 and 384, at two generations per ring, with heading $345.524\dg$ on a radius-384
flight against the compass spoke $345.5225\dg$, hat A's lane rotated by $120\dg$.

Its certificate closes with 134 situations and 449 transitions, no death, orientation locked, fewer
situations than any object of Table~\ref{tab:certificates}; the keyed flight has no generation off the certificate; and every corridor
quantity is hat A's: the reference cycle of 26 generations and $55.749$ units, the speed bound
$2.1396$, the heading interval of width $0.1189\dg$, and return words of 26, 42 and 68 generations with
displacements $A$, $A+B$ and $2A+B$, consecutive terms of the additive sequence one level below hat
A's, with count-weighted heading $345.522380\dg$ against hat A's $225.522493\dg+120\dg$
(Table~\ref{tab:outside}).

\subsection{One rule on both tilings: the spectre rule on the hat}
\label{sec:s33hat}

Gabbay's catalogue also contains a three-state, two-row hat traveller which he recognizes as our
spectre rule s33 under a relabelling of states. We had never run s33 on the hat. Unchanged, from
seeds of a few cells, it produces single gliders of five to six cells, loopers confined to a few rings
for thousands of generations, and showers of dozens of
travellers. A single glider keeps 5 to 6 cells from radius 48 to 384 and flies at $225.512\dg$ on a
radius-384 flight, within the error of a line fitted to its path of hat A's own spoke. Its certificate closes with 380
situations and $1{,}072$ transitions, no death; its cocycle is identical to hat A's; and its
return words are 42 and 68 generations with displacements $A+B$ and $2A+B$, count-weighted heading
$225.522425\dg$. On the spectre the same table flies the $48.014\dg$ lane family with the spectre's cocycle
($2{,}542$ situations, speed bound $1.5807$, heading interval $0.3926\dg$ wide); on the hat it flies the
$45.523\dg$ family with hat A's.

\begin{table}[t]
\caption{The three gliders from outside the census against hat A. Each certificate is its own, closed
from its own flight at the same $R$ and $k$; the return words are those of each object's most-visited
situation. The wake traveller's lane is hat A's rotated by $120\dg$.}
\label{tab:outside}
\begin{center}\small
\setlength{\tabcolsep}{2.5pt}\scriptsize
\begin{tabular}{@{}lllll@{}}
\hline
 & hat A & worm (2023) & wake (2026) & s33 on the hat\\
\hline
situations / transitions & 381 / 1{,}089 & 326 / 644 & 134 / 449 & 380 / 1{,}072\\
population in flight & 6--8 & 6--8 & 4--8 & 5--6\\
reference cycle (gen, u) & 26, 55.749 & 26, 55.749 & 26, 55.749 & 26, 55.749\\
speed bound (u/gen) & 2.1396 & 2.1396 & 2.1396 & 2.1396\\
heading interval width & $0.1189\dg$ & $0.1192\dg$ & $0.1189\dg$ & $0.1189\dg$\\
return times (gen) & 42, 68, 110 & 68, 110 & 26, 42, 68 & 42, 68\\
count-weighted heading & $225.522493\dg$ & $225.522495\dg$ & $345.522380\dg$ & $225.522425\dg$\\
\hline
\end{tabular}
\end{center}
\end{table}

\section{Discussion}
\label{sec:discussion}

\subsection*{The shared corridor}

Hat A and hat B, under different rules and with no situation in common, have the same number of
situations and identical cocycle numbers, but they are the same mechanism in two state alphabets
(Section~\ref{sec:seven}), so they show only that the search found that mechanism twice. The evidence that the corridor is set by the tiling comes from the objects
that are not relabellings of one another: the worm, the wake traveller, the spectre rule on the hat
and hat C's $105\dg$ glider, whose situation counts differ (326, 134, 380 and 257 against 381) but
whose cocycle numbers are the same. The corridor is fixed by the tiling and sets the speed, the
heading interval and the rhythm; a rule contributes the states it carries along it. A later search
of ours found the spectre's lane a second time, with a different rule. Trevorrow's worm
(Section~\ref{sec:worm}), under a rule designed three years before the corridor was found, and
designed only to run along the gaps between rows of reflected hats, carries its own states through
hat A's corridor with exactly hat A's cocycle numbers. Gabbay's traveller, independently evolved without a preferred direction,
carries the same numbers; so does our spectre rule transplanted to the hat. Why the corridor has these numbers, and why its return times form
the additive sequence they do, is answered in a paper in preparation: the tiles at which
the glider returns to a given situation form a row of the tiling's cut-and-project set, and the
return-word sequence is the coding of an irrational rotation.
Goucher's glider on the Penrose rhombs \cite{goucher} rides a straight ribbon of the tiling
\cite{debruijn}, and its immortality is a theorem of that geometry: progress along a ribbon is
monotone, so the path can never close. The monotilings have no ribbons, and here the certificate
does the work that the ribbon's geometry does there.

\subsection*{Engineering}

The earlier paper observed that aperiodicity removes the economics of construction: a component
verified at one site works elsewhere only at locally isomorphic recurrences of its exact environment,
which must be found rather than reached by translation. The certificate restores part of that
economy. A certified object is a component that needs verifying only once: its entire future at
any site whose situation lies in its closed set is Theorem~\ref{thm:immortality}, and the closed set
lists exactly those situations. The theorem is stated for a lone object, but a step depends only on the
collar, so it holds equally while every other object stays outside the collar. The parts a signalling
scheme would need are all present in the census: gliders carrying a passenger (a few cells that are not
themselves a glider), bound pairs that dissociate on a schedule, formation pairs (two gliders
travelling together), and the interaction repertoire of the earlier paper; a certificate
would give each of them an exact domain of validity.

\subsection*{Limitations}

Theorem~\ref{thm:immortality} is per object and per rule, and it certifies settled objects only; the
launch transient before the bridge generation is a finite recorded replay, not part of the theorem. Every computation rests on the semantics
of the substitution transducer of Ref.~\cite{tatham}: its soundness (T1) and its structural properties
(T3), (T4) follow from that paper's theorems and construction, while its completeness (T2) is checked
there algorithmically rather than proved. (T2) is supported here by exact geometry, by the faithfulness
check, by the patch comparisons, and by a cross-check against Tatham's recursive neighbor algorithm on
every legal edge address of at most twelve symbols on the hat and eleven on the spectre
(Appendix~\ref{app:crosscheck}).

\subsection*{Method}

Every control in this paper was stated before its result was known, and each
correction to the earlier paper comes from the certificates themselves.

\section{Reproducibility}
\label{sec:repro}

Everything reported here regenerates from committed artifacts; Appendix~\ref{app:repro} lists, claim by
claim, the artifact and the command. The certificate machinery is a single
program with subcommands for the atlas and its dump, the faithfulness check, the closure, the exact
cocycle and the keyed flights; the independent checker is a separate program in another language, which for the stratified Worms
rule is told which base type is the anti-hat.
Each certificate is a self-contained file: situations as keys, transitions with their exact motions,
the witness address of every ball used, and the launch bridge. Each keyed trace is a text file with
the command that produced it. The rules from outside the census are transcribed as data: HatLife's
Worms rule as an example program whose self-check steps the table rule and a direct transcription of
the transition function side by side before writing the record, and the other two as rule files read
by a generic runner. The code and the artifacts are at
\href{https://github.com/jamesedmond/monotile-ca}{github.com/jamesedmond/monotile-ca} (release 1.2.0)
and archived at Zenodo (doi:10.5281/zenodo.24178672; all versions doi:10.5281/zenodo.22836247), and the
earlier paper's interactive essay at
\href{https://offlattice.org/monotile/essay/}{offlattice.org/monotile/essay/}, which runs the same
engine, replays the gliders; its page \emph{Towards a proof} walks through situations, saturation and
closure. The census objects of Section~\ref{sec:certificate-scope} fly in live viewers at
\href{https://offlattice.org/monotile/census/}{offlattice.org/monotile/census/}.

\section*{Author's background}
The author is a software engineer working outside academia. This project began as a personal
investigation of cellular automata on the hat and spectre tilings and produced results that appear to
be new: the gliders of the earlier paper and, here, proofs that they travel forever
(Theorems~\ref{thm:immortality} and~\ref{thm:lane}). The work was carried out with substantial
assistance from an AI agent (Claude, Anthropic), which implemented the toolchain and searches,
drafted analyses, and drafted much of the text. The author directed the work throughout, and the
paper took its present form over many rounds of the author's line-by-line reading and discussion of
every section. The results rest on computations that the reader can rerun: each certificate is
re-derived by an independent program, and each control was stated before its result was known
(Sections~\ref{sec:seven} and~\ref{sec:repro}). The reader should expect correct and checkable
mathematics, but not necessarily the polish of a practised academic writer.

\section*{Acknowledgments}
Simon Tatham's finite-state transducers are the foundation of every patch and every atlas here.
Andrew Trevorrow pointed out his 2023 HatLife and corrected the earlier paper's priority claim;
Murdoch J. Gabbay's Aperiodic Life supplied the wake traveller and the observation that led to
Section~\ref{sec:s33hat}; both rules are used with their authors' permission. The author would also like
to thank Murdoch J. Gabbay, who read an earlier draft closely and whose suggestions shaped the
introduction and the note on the author's background.

\appendix

\section{The rules}
\label{app:rules}

The four automata of the earlier paper, Trevorrow's Worms and Gabbay's wake rule, printed from
their records (two redundant conjunctions in hat A's table are simplified, as noted below). Rows
are tried in order and the first matching clause fires; a cell matching no clause becomes quiescent. Clauses that a verified replay never uses
are retained as recorded. Hat A and B share the root
$(\text{anti-hat})(3,H_{0})(1,F_{0})\overline{(4,F_{0})}$, the overline marking the repeating part; the
spectre rules share the root
$(\text{spectre})(0,\Delta)\overline{(2,\Delta)}$.

\begin{center}
\small
\begin{tabular}{ccc}
\multicolumn{3}{l}{\textbf{Hat A (s21)}}\\
\hline
own & condition & next\\
\hline
0 & $n_{2}\ge2$ & 0\\
any & $n_{1}\ge3$ & 3\\
2 & $n_{1}\ge1$ and $n_{0}\ge1$ & 1\\
1 & $n_{2}\ge1$ & 3\\
2 & always & 0\\
0 & $n_{1}\ge1$ & 2\\
2 & $n_{3}\ge2$ & 0\\
2 & always & 0\\
\hline
\end{tabular}
\qquad
\begin{tabular}{ccc}
\multicolumn{3}{l}{\textbf{Hat B (s22)}}\\
\hline
own & condition & next\\
\hline
0 & $n_{1}\ge2$ and $n_{0}\ge1$ & 0\\
0 & $n_{0}\ge2$ and $n_{2}\ge2$ & 1\\
0 & $n_{0}\ge2$ and $n_{3}\ge1$ & 1\\
1 & $n_{3}\ge1$ & 3\\
\hline
\end{tabular}
\qquad
\begin{tabular}{ccc}
\multicolumn{3}{l}{\textbf{Hat C (s23)}}\\
\hline
own & condition & next\\
\hline
any & $n_{2}\ge1$ and $n_{1}\ge3$ & 3\\
1 & $n_{2}\ge1$ & 2\\
3 & $n_{3}\ge3$ & 3\\
0 & $n_{3}\ge2$ & 1\\
any & $n_{2}\ge1$ & 1\\
1 & $n_{0}\ge3$ and $n_{2}\ge2$ & 2\\
0 & $n_{3}\ge3$ & 3\\
\hline
\end{tabular}

\medskip
\begin{tabular}{ccc}
\multicolumn{3}{l}{\textbf{Spectre (s33)}}\\
\hline
own & condition & next\\
\hline
0 & $n_{0}\ge3$ and $n_{1}\ge1$ & 3\\
0 & always & 0\\
3 & $n_{3}\ge3$ & 0\\
0 & $n_{0}\ge3$ and $n_{1}\ge1$ & 0\\
3 & $n_{1}\ge1$ & 1\\
0 & always & 0\\
\hline
\end{tabular}
\qquad
\begin{tabular}{ccc}
\multicolumn{3}{l}{\textbf{Gabbay's wake rule (Aperiodic Life, 2026)}}\\
\hline
own & condition & next\\
\hline
1 & $n_{0}\ge3$ and $n_{2}\ge1$ & 2\\
0 & $n_{0}\ge4$ and $n_{2}\ge1$ & 1\\
1 & $n_{0}\ge4$ & 3\\
0 & $n_{1}\ge1$ and $n_{2}\ge1$ & 3\\
\hline
\end{tabular}

\medskip
\begin{tabular}{ccc}
\multicolumn{3}{l}{\textbf{Trevorrow's Worms (HatLife, 2023), hats}}\\
\hline
own & condition & next\\
\hline
0 & $n_{2}\ge2$ & 0\\
0 & $n_{1}\ge1$ & 0\\
0 & $n_{2}\ge1$ & 3\\
1 & $n_{1}\ge3$ & 0\\
1 & $n_{1}\ge2$ & 1\\
1 & $n_{1}\ge1$ and $n_{2}\ge2$ and $n_{3}\ge2$ & 0\\
1 & $n_{1}\ge1$ and $n_{2}\ge2$ and $n_{3}\ge1$ & 1\\
1 & $n_{1}\ge1$ and $n_{2}\ge2$ & 3\\
1 & $n_{1}\ge1$ and $n_{2}\ge1$ & 1\\
1 & $n_{1}\ge1$ and $n_{3}\ge2$ & 0\\
1 & $n_{1}\ge1$ and $n_{3}\ge1$ & 1\\
1 & $n_{1}\ge1$ & 3\\
2 & $n_{1}\ge2$ & 0\\
2 & $n_{1}\ge1$ and $n_{3}\ge4$ & 2\\
2 & $n_{1}\ge1$ and $n_{3}\ge3$ & 1\\
2 & $n_{1}\ge1$ & 2\\
3 & $n_{3}\ge3$ & 0\\
3 & $n_{2}\ge2$ & 0\\
3 & $n_{3}\ge2$ and $n_{2}\ge1$ & 2\\
\hline
\end{tabular}
\end{center}

Worms is stratified by chirality: hats carry the table above, and anti-hats (reflected hats) carry a table
with no rows, so they are quiescent forever (HatLife kills them every generation; since no row reads $n_{0}$, a
permanently quiescent cell and a missing one are the same to its neighbors). The ordered rows encode
HatLife's exact equalities, a row with threshold $k+1$ preceding the row with threshold $k$. States: 0
empty, 1 body, 2 head, 3 tail. The record carries the whole soup as its initial state, on the hat root
of the census at radius 96.

Two rows of hat A's table were simplified from the record's redundant conjunctions ($n_{1}\ge3$ and
$n_{1}\ge1$; $n_{3}\ge2$ and $n_{3}\ge3$); every other row is as recorded.

\medskip
\noindent\textbf{The records.} Every run regenerates from a record of family, root address, radius,
neighborhood (vertex throughout), rule and initial states; the cells of a patch are numbered in
breadth-first order from the root tile, cell 0. The hat root $(\text{anti-hat})(3,H_{0})(1,F_{0})\overline{(4,F_{0})}$
and the spectre root $(\text{spectre})(0,\Delta)\overline{(2,\Delta)}$ are the defaults $H$ and $S$ below.

\begin{center}\footnotesize
\begin{tabular}{@{}lllll@{}}
\hline
object & rule & root & radius & seed (cell: state)\\
\hline
hat A & s21 & $H$ & 48 & 0:1, 1:2\\
hat B & s22 & $H$ & 48 & 0:3, 1:1\\
hat C, both objects & s23 & $H$ & 48 & 0:2, 1:1\\
captured object & s21 & $H$ & 48 & 267:2, 210:1\\
spectre glider & s33 & $S$ & 48 & 0:1, 1:3\\
spectre, second family & s33 & $S'$ & 48 & 0:1, 4:3\\
Trevorrow's worm & Worms & $H$ & 96 & a random soup, as recorded\\
Gabbay's wake traveller & wake & $H$ & 96 & a random soup, as recorded\\
s33 on the hat & s33 & $H$ & 96 & a random seed, as recorded\\
\hline
\end{tabular}
\end{center}

Here $S'=(\text{spectre})(0,\Gamma)(0,\Psi)(1,\Psi)(1,\Pi)(7,\Xi)(7,\Delta)\overline{(2,\Delta)}$, the census
seed site of the second family. The three soup records carry their whole initial state; the seeds of the earlier paper's objects are
the two-cell launches above, from which each certificate's launch bridge names the generation at which
the flight enters the closed set.

\section{The atlas exploration}
\label{app:exploration}

The atlas is computed by exploring addresses through the neighbor transducer. A configuration
consists of a set of stored tiles with their types, classes and resolved gluings, one live transducer
run for every edge of every stored tile whose neighbor is not yet identified, a set of \emph{copy
links} for runs that have accepted, and the type of the highest supertile so far constrained. The
exploration starts from a single tile of the base type and feeds it the symbols of a possible
address, one level at a time, trying every legal parent type and slot at each level. Feeding a symbol
advances every live run; a run that emits its first output symbol thereby names the neighbor's type
and the back-edge by which it attaches, which places the neighbor exactly, so the neighbor is stored
and its own runs are spawned; later emissions are delivered to the runs of that neighbor as they
occur; a run that reaches an accepting state has established that the two tiles share every higher
supertile, so it is replaced by a copy link along which each further symbol is duplicated. Tiles are
stored only within a fixed region of the plane around the centre, large enough to contain every
$R$-ball; runs whose neighbor is geometrically certain to lie outside it are never started. A
configuration is a leaf once every tile within $R$ rings of the centre has a known class and no
unresolved edge: its $R$-ball is then exact, and no further symbol can change it. Configurations are
memoized by a key consisting of everything just listed, and two configurations with equal keys are
merged.

The check of Proposition~\ref{prop:graph} recomputes every node's key from its address path alone,
with none of this bookkeeping, and compares.

\begin{table}[h]
\caption{Number of legal $R$-balls on the two tilings (both base types on the hat), the data of
Figure~\ref{fig:atlas}. Each entry is the number of leaves of the checked exploration graph; the
exploration reaches address level at most five on every row.}
\label{tab:atlas}
\begin{center}
\setlength{\tabcolsep}{3.5pt}\scriptsize
\begin{tabular}{crrrrrrrrrrr}
\hline
$R$ & 2 & 3 & 4 & 5 & 6 & 7 & 8 & 9 & 10 & 11 & 12\\
\hline
hat & 118 & 195 & 287 & 394 & 544 & 719 & 897 & 1{,}081 & 1{,}282 & 1{,}504 & 1{,}733\\
spectre & 89 & 165 & 255 & 357 & 473 & 620 & 788 & 991 & 1{,}205 & 1{,}430 & 1{,}684\\
\hline
\end{tabular}
\end{center}
\end{table}

\section{The transducer against its reference algorithm}
\label{app:crosscheck}

To test (T2) directly, Tatham's recursive neighbor algorithm (Ref.~\cite{tatham}, Algorithm 1) was
reimplemented from its published pseudocode using only the library's public combinatorial tables (the
subtile lists, sub-edge counts and adjacency maps of each level of the hierarchy) and compared with the shipped
transducer on every legal edge address of increasing length, extending only the addresses that remain
undetermined, as that paper's corpus construction does. For each address the neighbor's address, the
neighbor's edge and the level of the lowest common supertile were compared with the transducer's
output and accept point. The adjacency maps were also checked to be fixed-point-free involutions
covering every interior and outline sub-edge symbol.

\begin{center}\small
\scriptsize
\begin{tabular}{@{}p{0.52\linewidth}rr@{}}
\hline
 & hat & spectre\\
\hline
addresses checked, of length at most & 22{,}388{,}882 (12) & 25{,}409{,}057 (11)\\
reference algorithm resolves; neighbor, edge and level agree & 9{,}306{,}354 & 11{,}462{,}096\\
both undetermined, extended & 12{,}034{,}197 & 13{,}517{,}116\\
transducer accepts where the reference algorithm overflows & 282{,}552 & 137{,}624\\
their continuations the algorithm resolves, all agreeing & 635{,}586 & 292{,}221\\
their continuations still undetermined at the length limit & 130{,}193 & 0\\
contradictions & 0 & 0\\
\hline
\end{tabular}
\end{center}

The early acceptances are genuine: they occur where the tile across an edge lies inside the same
supertile but the recursive algorithm can only reach it by climbing out across a zero-thickness spur
and back, which needs more symbols (Ref.~\cite{tatham}, Section 2.3); every continuation the
algorithm could resolve confirmed the transducer's answer.

In sum, the transducer and the reference algorithm agree wherever both decide, on 9{,}941{,}940 hat
addresses and 11{,}754{,}317 spectre addresses, with no contradiction; the check is of the transducer
against the combinatorial data, not against geometry. What it leaves open is set by its length limit
alone: the addresses the reference algorithm had still not decided at the limit, 302{,}446 of the
22{,}388{,}882 on the hat (1.4\%: the 172{,}253 early acceptances not followed up, counted in the
early-acceptance row, and the 130{,}193 continuations that ran out of symbols) and 105{,}155 of the
25{,}409{,}057 on the spectre (0.4\%, all early acceptances not followed up). These are the addresses
whose lowest common supertile lies above the level that twelve or eleven symbols reach, so a failure
of (T2) would have to be confined to supertile levels above every level the check reached, after
agreement at every level below. The check narrows where (T2) could fail without closing it, which is
the standing the main text gives it; for the balls and patches the certificates actually use, the
exact geometry, the faithfulness check and the patch comparisons listed in
Section~\ref{sec:discussion} bear on the same assumption. The program is a stand-alone crate in the
repository; it runs in about two minutes.

\section{Reproduction}
\label{app:repro}

Commands run from the repository root at release 1.2.0 after \texttt{cargo build --release}, which
puts \texttt{certify} and the examples in the release target directory. Each certificate file
records, under \texttt{entry}, the
launch it was closed from: record file and index, root, radius, start generation and component.
Angle brackets mark arguments.

\newcommand{\cmds}[1]{\par\smallskip\begingroup\small\ttfamily\raggedright\leftskip=1em\noindent #1\par\endgroup\smallskip}

\emph{Atlases} (Section~\ref{sec:atlas}, Table~\ref{tab:atlas}); \texttt{--verify} recomputes every
leaf's key from its path; likewise \texttt{spectre 9}, \texttt{hat 10}, \texttt{hat 11},
\texttt{spectre 11}. For Proposition~\ref{prop:graph}, \texttt{--dump} writes the exploration graph
per root type beside the atlas and \texttt{check-graph} re-derives every node. Validation against
patches, both ways, is the third command.
\cmds{certify atlas hat 9 --verify --out out/atlas-hat-r9.json\\
certify atlas hat 10 --verify --dump --out out/atlas-hat-r10.json\\
\hspace*{1em}certify check-graph hat <graph file>\\
certify validate hat out/atlas-hat-r9.json --radius 128}

\emph{Certificates} (Table~\ref{tab:certificates}): the closure with the arguments the file's
\texttt{entry} records; the cocycle of Theorem~\ref{thm:lane}; the class-blind count, branching
and connectivity.
\cmds{certify closure <record> <atlas> --radius 384\\
\hspace*{1em}--start <generation> --k 4 [--index <i>]\\
\hspace*{1em}[--component farthest] --out <file>\\
certify cocycle <file>\\
certify stats <file>}

\emph{Controls} of Section~\ref{sec:seven}: death, by closing hat A's launch under hat B's rule;
seeding, by closing from a later generation and comparing; collar radius, by closing at
\texttt{--k 2}, \texttt{5} or \texttt{6} with the $R=12$ atlases and projecting the finer closure
onto the coarser collar.
\cmds{certify closure <hat A record> <atlas> ... \\
\hspace*{1em}--rule-from results/hat-tableevolve-r48-s22.jsonl\\
certify closure <record> <atlas> ... --seed-gen 300\\
\hspace*{1em}--compare <certificate>\\
certify stats <fine closure> <coarse closure> --project 4}

\emph{Independent checker}; the stratified Worms certificate takes \texttt{--strata 1:1}.
\cmds{python3 crates/certify/check/check\_certificate.py\\
\hspace*{1em}<certificate> <atlas> <record> [--index <i>]}

\emph{Keyed flights} (control 1; Section~\ref{sec:outside}).
\cmds{certify trace <record> <certificate> --gens 20000\\
\hspace*{1em}--launch-gens <g> --launch-radius <r>\\
\hspace*{1em}[--select-heading <degrees>] --out <tsv>}

\emph{Flights and Figure~\ref{fig:certcompass}}: the flights are \path{results/flights/*.csv}; the
figure is drawn by \path{crates/certify/figures/fig_certcompass.py}.

\emph{Gliders from outside the census} (Section~\ref{sec:outside}): the rules as data in
\path{results/rules/}, run from random soups; soups 8 and 14 of the first command give the
committed wake and s33-on-hat records, and the Worms rule's own example, which checks the
transcription against a direct implementation of HatLife's rule, gives the worm's record; the
reflected-hat rows of Section~\ref{sec:worm} come from the parametric table of the radius-512 patch.
\cmds{tiling-core --example table\_soups <rule.json>\\
\hspace*{1em}hat 96 400 16 50 24301 2 8\\
\hspace*{1em}results/hat-wake-gabbay-soup8.jsonl\\
tiling-core --example table\_soups <rule.json>\\
\hspace*{1em}hat 96 400 16 50 24301 2 14\\
\hspace*{1em}results/hat-s33-soup14.jsonl\\
tiling-core --example worms\_hatlife\\
\hspace*{1em}(record results/hat-worms-trevorrow-soup2.jsonl)\\
certify param hat '<root>' 512 --out <tsv>\\
python3 crates/certify/figures/antihat\_rows.py <tsv>}

\emph{Hat A and hat B} (Section~\ref{sec:seven}).
\cmds{certify relabel <A certificate> <B certificate>\\
\hspace*{1em}--map 1:3,2:1,3:0\\
tiling-core --example track\_compare <A record>\\
\hspace*{1em}<B record> 256 440 225 60 4}

\emph{Census} (Section~\ref{sec:certificate-scope}); its tables are under
\path{results/certificates/census3/}, its seeds in \path{census-seeds.jsonl}. For each of its objects,
\path{results/certificates/census-certificates.tsv} lists the record, the closure command that rebuilds
the certificate, and the resulting situation count, heading window and speed bound; the second command
rebuilds every one and compares it with that table.
\cmds{certify census <record> <atlas> --radius 384\\
python3 crates/certify/census/rebuild\_certificates.py}

\emph{Transducer cross-check} (Appendix~\ref{app:crosscheck}): a stand-alone crate in the check
directory, run on its own.
\cmds{cd crates/certify/check/transducer-crosscheck\\
cargo run --release}

\emph{Figures}: the scripts in the certify crate's figures directory (the \texttt{fig} scripts
and \texttt{portraits.sh}) read the dumps of the two commands below.
\cmds{tiling-core --example hier\_dump <record> 0 <radius> <out.json>\\
certify situation-dump <certificate> <record>\\
\hspace*{1em}--state <s> --balls <list> --out <out.json>}

\end{document}